\documentclass[hyper]{prop2015}
\usepackage[english]{babel}

\usepackage{bbm,amsmath,amssymb}   
\usepackage{epsfig}                    
\usepackage[matrix,arrow]{xy}          
\usepackage{xspace}                    
\usepackage{stmaryrd}                  
\usepackage{slashed}
\usepackage{appendix,graphicx}
\usepackage{enumitem}

\usepackage{units} 
\usepackage{tikz}
\usepackage{color}
\newcommand{\remark}[1]{}

\newcommand{\dd}{\mathrm{d}}
\newcommand{\ee}{\mathrm{e}}
\newcommand{\ii}{\mathrm{i}}

\newcommand{\der}{\partial}

\newcommand{\bbR}{\mathbbm{R}}
\newcommand{\bbC}{\mathbbm{C}}

\DeclareMathOperator{\SU}{\mathit{SU}}
\DeclareMathOperator{\SO}{\mathit{SO}}
\DeclareMathOperator{\SL}{\mathit{SL}}
\DeclareMathOperator{\GL}{\mathit{GL}}
\DeclareMathOperator{\Symp}{\mathit{Sp}}
\DeclareMathOperator{\Spin}{\mathit{Spin}}

\newcommand{\rep}[1]{\mathbf{#1}}

\newcommand{\id}{\mathbbm{1}}

\DeclareMathOperator{\tr}{tr}

\newcommand{\Lgen}{L}

\newcommand{\Dgen}{{D}}

\DeclareMathOperator{\adj}{ad}

\newcommand{\proj}[1]{\times_{#1}}

\DeclareMathOperator{\Edd}{\mathit{E_{d(d)}}}

\newcommand{\ra}{\rightarrow}

\newcommand{\cN}{\mathcal{N}}

\newcommand{\GenStr}{\mathcal{G}}
\newcommand{\GenHd}{\mathcal{H}}

\DeclareMathOperator{\End}{\text{End}}
\def\Q{\mathcal{Q}}

\category{Proceedings}
\subtitle{\href{http://www.maths.dur.ac.uk/lms/109/index.html}{LMS/EPSRC Durham Symposium on Higher Structures in M theory}}
\title{Supergravity Fluxes and Generalised Geometry}
    \author[C. Strickland-Constable]{Charles Strickland-Constable\inst{a,}\footnote{Corresponding author e-mail:~\href{mailto:c.strickland-constable@herts.ac.uk}{\textsf{c.strickland-constable@herts. ac.uk}}}}
\address[1]{School of Physics, Astronomy and Mathematics, University of Hertfordshire, College Lane, Hatfield, AL10 9AB, United Kingdom}
\shortauthors{C. Strickland-Constable}
\begin{abstract}
  We briefly review the description of the internal sector of supergravity theories in the language of generalised geometry and how this gives rise to a description of supersymmetric backgrounds as integrable geometric structures. 
  We then review recent work, featuring holomorphic Courant algebroids, on the description of $\cN=1$ heterotic flux vacua. This work studied the finite deformation problem of the Hull--Strominger system, guided by consideration of the superpotential functional on the relevant space of geometries. It rewrote the system in terms of the Maurer--Cartan set of a particular $L_\infty$-algebra associated to a holomorphic Courant algebroid, with the superpotential itself becoming an analogue of a holomorphic Chern--Simons functional.
\end{abstract}
\shortabstract
\begin{document}
\maketitle


\section{Introduction}


In this contribution, we will briefly summarise recent results in the broad area of supersymmetric solutions and compactification of the supergravity limits of string theories and M-theory. This is a subject with a long history, and we will not attempt to give a full discussion of it here (see~\cite{Grana:2005jc} for a review). 
Our purpose will be to describe some specific developments over the last few years. 
In particular, we will examine how the notion of generalised geometry, as introduced in~\cite{Hitchin:2004ut,Gualtieri:2003dx}, together with its exceptional~\cite{Hull:2007zu,Pacheco:2008ps} and heterotic~\cite{Garcia-Fernandez:2013gja,Coimbra:2014qaa} extensions, has provided useful insights. 

In Section~\ref{sec:Gen-geom}, we give a schematic description of how it provides a geometrical formulation of the internal sectors of maximal ten- and eleven-dimensional supergravities, in a way that unifies their bosonic fields and symmetries. We mainly follow~\cite{Coimbra:2011nw,Coimbra:2011ky,Coimbra:2012af} and refer the reader to them for both details of the construction and a more complete bibliography. 
We then give an overview of how this gives rise to an elegant characterisation of their supersymmetric Minkowski backgrounds 
as integrable geometric structures.
Then, in Section~\ref{sec:heterotic}, 
we review how the holomorphic bundle structures introduced in the study of infinitesimal heterotic moduli~\cite{Anderson:2014xha,delaOssa:2014cia} carry the structures of holomorphic Courant algebroids, and make contact with the generalised geometry picture. 
These algebroid structures further give rise to $L_\infty$-algebras and the superpotential functional becomes an extension of the holomorphic Chern--Simons functional which appears to be a natural object associated to them.
A separate article to be published in this volume will include comments on some of the open problems and future goals of these programmes. 


\section{Generalised geometry, supersymmetric backgrounds and consistent truncations}
\label{sec:Gen-geom}


Before we start to discuss generalised geometry, let us note that we will mostly be discussing not the full ten- and eleven-dimensional supergravity theories, but only some internal sector. These sectors are defined by imposing a warped metric ansatz on the theory of the type
\begin{equation}
	\hat{g} = \ee^{2\Delta(x)} \eta_{\mu \nu} \dd y^\mu \dd y^\nu + g_{mn}(x) \dd x^m \dd x^n
\end{equation}
where $\eta_{\mu \nu} \dd y^\mu \dd y^\nu$ will be the external (warped) Minkowski (or possibly AdS) metric, and $\Delta$ is a scalar warp factor depending only on the coordinates of the internal space. We then truncate the fields of the theory, keeping only degrees of freedom which are scalars with respect to the external Lorentz symmetry. This defines a Euclidean theory living on the internal space, whose (supersymmetric) solutions are (supersymmetric) Minkowski (or AdS) vacua of the full ten- or eleven-dimensional theory. We refer the reader to~\cite{Coimbra:2012af} for a detailed example of how this works in the case of eleven-dimensional supergravity. 
It will be precisely these internal sectors which will become geometrical when we study them in the language of generalised geometry, and when we refer to supergravity in this section, we are often meaning only this internal sector. 
The full theory can be formulated in a language where the internal sector is described by this construction, see e.g.~\cite{Godazgar:2014nqa} in the context of exceptional field theory. 
For the purposes of our discussion here, we will restrict to the cases where the external space has dimension at least four, though see~\cite{Hohm:2014fxa,Bossard:2018utw} for recent studies which go beyond this.


\subsection{Supergravity as generalised geometry}

One of the main ideas underlying the formulation of supergravity in terms of generalised geometry is the unification of symmetries. A key symmetry of gravitational theories is provided by the diffeomorphism group of spacetime. In supergravity, one typically has additional $p$-form fields in the bosonic field content, which carry additional gauge symmetries. Generalised vectors unify these symmetries by combining a vector field, which generates a diffeomorphism, with other differential forms (or sometimes also more general tensor fields) which generate the gauge transformations of the remaining physical fields. The combined transformations are often referred to as generalised diffeomorphisms. 

For example, in the NS-NS sector of type II supergravity, one has a two-form field $B$, which undergoes shift symmetries $\delta B = \dd \lambda$. As such, the generalised vector in this case is composed of a vector $v$ together with a one-form $\lambda$, so that $V = v+\lambda$ is a section of the generalised tangent bundle 
\begin{equation}
	E \simeq TM \oplus T^*M .
\end{equation}
In fact, due to the local nature of $B$, this representation is also local, with the full picture involving transition functions due to gauge transformations. This is often referred to in the literature as twisting by the $B$ field gerbe. We refer the reader to~\cite{Coimbra:2011nw} for a full discussion.

More generally, the generalised tangent bundle takes the form $E\simeq TM \oplus \dots$ 
where $\dots$ denotes a sum of 
additional differential form parts, and sometimes also more general tensors. 
The gauge transformations which patch it together live in a parabolic subgroup of some real Lie group containing $\GL(d,\bbR)$,
where $d$ is the dimension of the internal space on which we are working. 
This Lie group, here denoted $\GenStr$, is the generalised structure group 
and it is determined by the physical theory under consideration. 
For example, for the NS-NS sector example above it is $O(d,d)\times\bbR^+$, while for eleven-dimensional supergravity on a $d$ dimensional internal space it is $\Edd\times\bbR^+$. These are of course the continuous forms of the relevant T duality and U duality groups that appear in compactifications on a torus. 
The fibre of the generalised tangent space forms a representation of $\GenStr$, whose Dynkin label follows a schematic pattern described in~\cite{Strickland-Constable:2013xta} (see also~\cite{Cederwall:2017fjm} for further discussion). 

One can introduce a generalised metric as a positive definite inner product on the generalised tangent space which breaks the structure group $\GenStr$ to its maximal compact subgroup, here denoted $\GenHd$. This corresponds precisely to the data of a physical field configuration of the theory, unifying the metric and the supergravity gauge fields. 


Having defined the generalised tangent bundle $E$, one then finds that many of the usual concepts of differential geometry carry over to the generalised geometry construction. A very important example is the Lie derivative, which provides the action of infinitesimal diffeomorphisms on tensor fields. In generalised geometry, this becomes the Dorfman derivative which gives the action of infinitesimal generalised diffeomorphisms on generalised tensor fields. It follows a universal formulation~\cite{Coimbra:2011ky}
\begin{equation}\label{eq:Dorfman}
	\Lgen_V = \der_V - (\der \proj{\adj} V) \cdot ,
\end{equation}
where the second term denotes the adjoint action of the partial derivative of the generalised vector $V$ projected onto the adjoint bundle of the generalised structure group $\GenStr$. Decomposing this into ordinary vectors and differential forms it is straightforward to see that it becomes the Lie derivative along the vector part of $V$ together with the appropriate adjoint actions of the exterior derivatives of the form parts of $V$. Acting on the generalised metric, it provides the infinitesimal transformations of the physical fields under the bosonic symmetries when one decomposes into conventional objects. 
Acting on generalised vectors, one can anti-symmetrise the Dorfman derivative to obtain the Courant bracket. 

One can also define generalised connections to be linear differential operators acting on generalised tensor bundles $Y$ (i.e. vector bundles whose fibres are representations of $\GenStr$ and whose transition functions match those of $E$)
\begin{equation}\label{eq:Gen-con}
\begin{aligned}
	\Dgen : Y &\longrightarrow E^* \otimes Y , 
\end{aligned}
\end{equation}
which act only in the Lie algebra of $\GenStr$, so that they preserve invariant tensors of $\GenStr$. Using the Dorfman derivative one can then define a natural notion of generalised torsion, exactly analogous to the usual definition in terms of the Lie derivative. One can then go on to study torsion-free connections which preserve a given generalised metric: the analogues of the Levi-Civita connections familiar from ordinary geometry and general relativity. One finds that such connections always exist, but contrary to the usual situation, they are not uniquely determined by the generalised metric. However, this does not turn out to be of much concern for the formulation of supergravity equations; every time we wish to write such an equation, the undetermined parts of the generalised connection will cancel out.

These objects are essentially all we need to formulate the equations of our supergravity theory~\cite{Coimbra:2011nw,Coimbra:2011ky,Coimbra:2012af}. The supersymmetry parameters and fermionic fields can be viewed as sections of vector bundles transforming under the maximal compact subgroup $\GenHd$ of the generalised structure group. We can then build operators acting on these objects, by acting with a generalised Levi-Civita connection $D$ and subsequently projecting that object ($\GenHd$-covariantly) onto one of its $\GenHd$-irreducible parts. For example, one can write the supersymmetry variation of the gravitino $\psi$ by the supersymmetry parameter $\epsilon$ schematically as
\begin{equation}\label{eq:Gravitino-SUSY}
	\delta \psi = \Dgen \times \epsilon ,
\end{equation}
where the symbol $\times$ here denotes projection onto the bundle of which the gravitino field is a section. As stated above, the operator appearing in~\eqref{eq:Gravitino-SUSY} will be uniquely determined by the generalised metric, and thus independent of the choice of generalised Levi-Civita connection.

Such operators provide generalised geometry expressions encapsulating the supersymmetry variations of the fermionic fields. 
Similar operators acting on the fermionic fields can be used to write the fermionic equations of motion. 
Subsequently one can consider the supersymmetry variations of those fermionic equations. Closure of the supersymmetry dictates that the result must be the bosonic equations of motion. Indeed one finds that there exists a generalised Ricci curvature tensor which arises precisely out of considering this closure, and that its vanishing is precisely the bosonic equations of motion. Finally, the supersymmetry variations of the bosons are simple $\GenHd$-covariant contractions of the fermionic fields with the supersymmetry parameters. 

Thus, we have a formulation of the equations of the theory in terms of generalised geometry objects. A major advantage of this formulation over the conventional equations is that it has manifest enhanced symmetry, 
which considerably simplifies calculational manipulations. 
For example, the usual statement for Minkowski backgrounds that the supersymmetry conditions (plus the Bianchi identities) imply the equations of motion becomes almost immediate in this language. 
This is directly analogous to the Ricci-flatness of Calabi--Yau or $G_2$ holonomy manifolds.


\subsection{Supersymmetric backgrounds}

Having formulated the theory in generalised geometry language, one would like to consider applications of the machinery. One problem for which it is especially appropriate is the classification of supersymmetric backgrounds with non-zero fluxes
\footnote{There are standard no-go theorems forbidding compact smooth solutions with fluxes~\cite{Maldacena:2000mw,Ivanov:2000fg} if the external space is Minkowski. In order to have such compactifications in string theory one needs to include negative tension sources: orientifold planes lying along some hypersurfaces in the space. In this article it is understood that in those cases, we are describing the geometry away from the locus of the sources.}. 
Again, we consider first the situation for these in terms of ordinary geometric structures.

One of the major tools that has been applied to the classification of supersymmetric backgrounds, starting with~\cite{Gauntlett:2002sc,Gauntlett:2002nw,Gauntlett:2002fz,Gauntlett:2003cy}, is the notion of a $G$-structure. A $G$-structure on the tangent bundle of a manifold is a covering of the manifold with open sets equipped with local frames $\hat{e}_a$ for the tangent bundle, such that on the overlaps of patches, the frames are related by transformations only in the subgroup $G \subset \GL(d,\bbR)$.

For example, given a Riemannian metric we can choose local orthonormal frames $\hat{e}_a$ on patches covering our manifold. As orthonormal frames are related by $\SO(d)$ transformations (assuming also an orientation), such frames define an $\SO(d)$ structure on the tangent bundle. 

Further reductions of the structure group are possible. For example, in seven-dimensions, one could have a non-vanishing spinor field $\epsilon$. Such a spinor would be stabilised by $G_2 \subset \SO(7)$ and frames $\hat{e}_a$ with respect to which only the first component of the spinor were non-zero, would be related by $G_2$ transformations. Thus the spinor defines a $G_2$ structure. It is useful to note that this structure could also be specified by a three-form $\phi$ lying in a particular (open) orbit of $\GL(7,\bbR)$, which would also be stabilised by $G_2 \subset \GL(7,\bbR)$. This would be related to the spinor via the identification with its three-form bilinear $\phi_{mnp} = \bar\epsilon \gamma_{mnp} \epsilon$. There are many cases in which a given $G$-structure can be defined either from the existence of preserved spinors or from the existence of preserved forms or tensors.  

Another example of a $G$-structure is provided by an almost complex structure on an even-dimensional manifold. The endomorphism $J \in \End{TM}$ (with $J^2 = -\id)$ splits the complexified tangent bundle into a sum of $\pm\ii$ eigenbundles. On our patches of the manifold we can thus choose a frame for the $+\ii$ eigenbundle and use the union with its complex conjugate as a frame for $TM$. Such frames are related by $\GL(n,\bbC) \subset \GL(2n,\bbR)$ transformations, where $2n$ is the dimension of the manifold, so we have a $\GL(n,\bbC)$ structure. This is an example of a ``non-metric" $G$-structure, as $\GL(n,\bbC)$ is not a subgroup of $\SO(2n)$. 

For any $G$-structure, there is an associated tensor known as its intrinsic torsion. Roughly, this is defined as follows (see e.g.~\cite{Joyce:2000yy} for a more detailed treatment). A tangent bundle connection is said to be compatible with a $G$-structure if it acts only in the Lie algebra of the structure group $G$, equivalently if it preserves all invariant tensors of the structure.
If one considers the torsion of such connections, one finds that a particular projection of the torsion is the same for all compatible connections, and it is generically non-zero. This projected object is called the intrinsic torsion of the $G$-structure and it is defined purely by the geometry of the structure. It's vanishing is equivalent to the existence of a torsion-free compatible connection. 
In this review, we will label $G$-structures for which the intrinsic torsion vanishes as integrable. 

What does this condition imply for our examples above? In the case of a metric it gives no additional condition, as we expect from the known existence of the Levi-Civita connection. For the $G_2$ structure, the condition can be expressed either in terms of the spinor $\epsilon$ or the three-form $\phi$. For the spinor, we have 
that it is parallel with respect to the Levi-Civita connection, so that this connection has $G_2$ (restricted) holonomy. Equivalently, we could impose that $\phi$ is both closed and co-closed. For the almost complex structure, vanishing intrinsic torsion is equivalent to the usual Nijenhuis tensor condition, so that the almost complex structure is integrable and we have a complex manifold.

Such $G$-structures can be used to characterise supersymmetric backgrounds~\cite{Gauntlett:2002sc,Gauntlett:2002nw,Gauntlett:2002fz,Gauntlett:2003cy}, where the Killing spinors on the internal space (which provide the supersymmetry) are stabilised by the group $G$. The differential equations that they satisfy specify the intrinsic torsion of the $G$-structure in terms of the fluxes present in the solution. This approach has been very useful in constructing and classifying solutions (see e.g.~\cite{Gauntlett:2002nw,Gauntlett:2002fz,Kaste:2003zd,Gauntlett:2004zh,Lukas:2004ip,Lust:2004ig,Gauntlett:2005ww}) but has some limitations. In particular, it is possible that the stabiliser group $G$ may not be the same at all points in the space, leading to interpolating cases which are more awkward to handle. Also, the structures are not torsion-free, which can lead to significant complications if one is interested studying further properties, such as their moduli. We will see below that considering these systems as generalised structures cures both of these issues.

A generalised $G$-structure is defined completely analogously to the above. One simply insists on the existence of local frames (in the generalised frame bundle, see~\cite{Coimbra:2011ky}) for the generalised tangent bundle which are related by $G$-transformations on the overlaps of patches. Also, exactly as above, there is a notion of generalised intrinsic torsion associated to these structures~\cite{Coimbra:2014uxa}, and we will describe structures for which this vanishes as integrable. 

In $\SO(6,6)\times\bbR^+$ generalised geometry, the two Killing spinors of an $\cN=2$ vacuum of type II supergravity define an $\SU(3)\times\SU(3)$ structure~\cite{Grana:2004bg}, 
which can equivalently be specified by a pair of polyforms $\Phi^\pm$ of even and odd degree respectively, constructed from their bilinears. These polyforms can be seen as pure spinors of $\SO(6,6)$. The conditions for supersymmetry (in the absence of RR-flux, which is not included as part of $\SO(6,6)\times\bbR^+$ generalised geometry, but can be added by hand into the equations) were seen~\cite{Grana:2004bg} to be $\dd \Phi^\pm = 0$, which is the condition for vanishing generalised intrinsic torsion of this structure. This integrability can also be seen as the existence of a generalised Levi-Civita connection with respect to which the Killing spinors are parallel. By analogy with the case of ordinary Riemannian geometry, where vanishing intrinsic torsion is equivalent to special (restricted) holonomy, this generalised geometry condition was called generalised special holonomy in~\cite{Coimbra:2014uxa}. The construction also avoids the problem of interpolating structure groups by associating the spinors with different $\SO(6)$ subgroups of $\SO(6,6)$. Viewing them in this way, the stabiliser of the pair is always $\SU(3)\times\SU(3)$ regardless of their relative orientation. 
The pure spinor form of the equations have been fruitfully applied to the classification and construction of solutions~\cite{Grana:2006kf,Andriot:2008va,Apruzzi:2013yva,Apruzzi:2014qva,Apruzzi:2015zna,Andriot:2015sia,Passias:2017yke}. 

\begin{table}[htb]
\centering
\begin{tabular}{llll}
   $d$ & $\GenStr$ & $\GenHd$ &  $G_{\cN}$ \\ 
   \hline
   4 & $E_{7(7)}\times\bbR^+$ & $\SU(8)$ &  $\SU(8-\mathcal{N})$\\
   5 & $E_{6(6)}\times\bbR^+$& $ \Symp(8)$ & $\Symp(8-2\mathcal{N})$ \\
   6 & $\Spin(5,5)\times\bbR^+$& $\Symp(4)\times \Symp(4)$ &   
      $\Symp(4-2\mathcal{N}_+)\times$\\
       & &  & $~~~~\times\Symp(4-2\mathcal{N}_-)$\\
   7 & $\SL(5,\bbR)\times\bbR^+$& $\Symp(4)$ & $\Symp(4-2\mathcal{N}) $
\end{tabular}
\caption{\label{tab:Gen-holonomy}Integrable generalised structure groups $G_{\cN}$ for $d$-dimensional Minkowski backgrounds of type II and eleven-dimensional supergravity preserving $\cN$ supersymmetries.}
\end{table}

The correspondence between supersymmetric Min\-kowski backgrounds and generalised special holonomy can be extended to include the RR fluxes (i.e. all internal fluxes) in $\Edd\times\bbR^+$ generalised geometry. The result is that Minkowski backgrounds in Type II theories or eleven-dimensional supergravity are precisely those spaces admitting integrable generalised $G$-structures with the structure groups listed in table~\ref{tab:Gen-holonomy}. These results were established by showing that the Killing spinor equations precisely set the various components of the generalised intrinsic torsion to zero~\cite{Coimbra:2014uxa,Coimbra:2016ydd}. Descriptions of the conditions in terms of bosonic bilinears of the Killing spinors in the 8-supercharge and half-maximal cases can be found also in~\cite{Ashmore:2015joa,Ashmore:2016qvs,Malek:2016bpu,Malek:2017njj}. 

This picture can be adapted to give a description of supersymmetric AdS backgrounds~\cite{Ashmore:2016qvs,Malek:2017njj,Coimbra:2015nha,Coimbra:2017fqv}. The modification is simply to switch on a singlet generalised intrinsic torsion\footnote{Technically, one must make an assumption about the inner products of the Killing spinors to guarantee the existence of this structure for even-dimensional AdS spaces~\cite{Coimbra:2017fqv}, though it is expected that the other solutions are simply AdS foliations of higher dimensional AdS geometries.}, making these structures analogous to Sasaki-Einstein manifolds or weak $G_2$ manifolds, rather than spaces with special holonomy. In both the Minkowski and AdS cases, the (weakly) intergrable generalised structure immediately provides the Killing superalgebra of the solutions~\cite{Coimbra:2016ydd,Coimbra:2017fqv}, providing a proof that these match the expected symmetries of the dual CFTs in the AdS case.
These ideas were also used to demonstrate the gravity dual of the statement that the exactly marginal deformations are the marginal deformations quotiented by the complexified global symmetries~\cite{Ashmore:2016oug}.


\section{Heterotic supergravity and deformations of $\cN=1$ backgrounds}
\label{sec:heterotic}


A generalised geometry description of heterotic supergravity has also been proposed in~\cite{Garcia-Fernandez:2013gja,Coimbra:2014qaa} based on the generalised tangent bundle
\begin{equation}
E \simeq  TM \oplus \End V \oplus \End \, TM \oplus  T^*M
\end{equation}
%
where $V$ is a vector bundle transforming under the heterotic gauge group. 
In this construction, a generalised Levi-Civita type connection is determined not via the condition that it is torsion-free, but via insisting that its associated operators and curvatures reproduce the relevant supergravity equations. Indeed, the connection does not satisfy the natural torsion-free condition with respect to the Dorfman derivative, and the connection between supersymmetric backgrounds and vanishing intrinsic torsion is lost.

However, it was found~\cite{delaOssa:2014cia,Anderson:2014xha} that the infinitesimal moduli of generic $\cN=1$ Minkowski backgrounds~\cite{Hull:1986kz,Hull:1986iu,Strominger:1986uh} are given by $H^1_{\bar{D}}(\Q)$ where the bundle 
\begin{equation}
\Q \simeq T^{(1,0)}X  \oplus \End V \oplus \End TX \oplus T^{*(1,0)}(X)
\end{equation}
has a natural holomorphic structure
\begin{equation}
\bar{D}\colon \Omega^{0,p}(\Q) \rightarrow  \Omega^{0,p+1}(\Q),
\end{equation}
built out of the Dolbeault operator on the complex manifold $X$ and the supergravity field strengths, which squares to zero on imposing the Bianchi identity for the fluxes. 
In light of this, 
it is natural to speculate that a generalised geometry based on this bundle could be a useful tool in studying these systems. 
In~\cite{Ashmore:2018ybe}, the finite deformations of the system were studied both as an interesting problem in their own right, and as a means to uncover the relevant generalised geometry of $\Q$.

The Hull--Strominger system formulates the conditions for a four-dimensional $\cN=1$ Minkowski background in terms of a 2-form $\omega$ and a 3-form $\Omega$ given by
\begin{equation}
\omega_{mn} = -\ii\,\eta^\dagger \gamma_{mn} \eta,
\qquad 
\Omega_{mnp} = \ee^{-2\phi} \eta^\text{T}\gamma_{mnp}\eta,
\end{equation}
where $\eta$ is a non-vanishing spinor field on the internal six-manifold $X$. Their statement is that these objects, together with the heterotic supergravity 3-form field strength  
\begin{equation}
\label{eq:het-H}
	H = \dd B + \frac{\alpha'}{4}(\omega_\text{CS}(A)-\omega_\text{CS}(\Theta))
\end{equation}
and the curvature $F$ of the gauge fields $A$ must satisfy the equations
\begin{align}
\dd \Omega &= 0, \label{eq:HS1}\\
\ii (\partial - \bar \partial) \omega& = H 
, \label{eq:HS2}\\
\Omega \wedge F &= 0, \label{eq:HS3}\\
\omega \lrcorner F &=  0, \label{eq:HS4}\\
\dd (\ee^{-2\phi} \omega \wedge \omega) &= 0, \label{eq:HS5}
\end{align}
together with the heterotic Bianchi identity
\begin{equation}
\dd H = \frac{\alpha'}{4}(\tr F \wedge F - \tr R \wedge R) \label{eq:HS6}.
\end{equation}
Note that equation~\eqref{eq:HS1} implies that the manifold $X$ is complex with holomorphically trivial canonical bundle, while~\eqref{eq:HS2} is the statement that the gauge bundle $V$ is holomorphic on $X$. 

Equations~\eqref{eq:HS1},~\eqref{eq:HS2} and~\eqref{eq:HS3} are often referred to as the F-term conditions, and we will focus on these conditions only in what follows as it is argued in~\cite{Ashmore:2018ybe} that their moduli space is in fact the full moduli space of the system. 
They can be derived from the superpotential functional~\cite{Gurrieri:2004dt}
\begin{equation}
W=\int_{X}(H+\ii\,\dd\omega)\wedge\Omega ,
\end{equation}
by imposing both its vanishing and that its naive variations with respect to the objects in its definition are also zero~\cite{delaOssa:2015maa}:
\begin{equation}\label{eq:super_first}
W = \delta W = 0
\end{equation}
A question which arises from this is: what exactly is the space of geometries on which we have defined the functional $W$? Further still, we would like to know how to describe some of its properties. In particular, considering the full (i.e. unreduced) ten-dimensional supergravity theory as a four-dimensional $\cN=1$ supergravity theory with an infinite number of fields (an approach pioneered in~\cite{deWit:1986mz} in the $\cN=8$ case and later exploited in the $\cN=2$ case in~\cite{Grana:2005ny,Grana:2009im,Ashmore:2015joa}), we see that this space of geometries should correspond to scalar fields in four-dimensions. Thus, as is standard in $\cN=1$ supergravity, it is should have a K\"ahler structure and in particular this means that it should be complex. 
Below, we will be interested to know 
how to describe the complex structure on this space.

To define the space of $\cN=1$ structures more carefully, we consider first the simpler case of an ordinary almost complex structure. These can be viewed either as an endomorphism of the tangent bundle $J$ which squares to minus the identity, or equivalently as 
a split of the complexified tangent bundle into the $\pm\ii$ eigenbundles of $J$
\begin{equation}
	TM_\bbC = T^{(1,0)} \oplus T^{(0,1)}
\end{equation}
One can specify a finitely deformed (but nearby) almost complex structure by specifying the new $-\ii$ eigenbundle as the graph of a map
\begin{equation}
	\mu : T^{(0,1)} \longrightarrow T^{(1,0)}
\end{equation}
(i.e. as the image of the map $\id + \mu$ acting on $T^{(0,1)}$). Such a map can also be thought of as a $(1,0)$-vector valued $(0,1)$-form
\begin{equation}
	\mu \in \Omega^{0,1}(T^{(1,0)})
\end{equation}
As such objects uniquely label nearby almost complex structures, we view the tensor fields $\mu$ as a set of (complex) coordinates on the (infinite-dimensional) space of almost complex structures in a neighbourhood of $J$. Note that if we were to write the new (real) almost complex structure $J'$ in terms of the original one $J$ and our coordinate $\mu$, we would get an infinite series of terms involving both $\mu$ and $\bar\mu$. 
However, taking a formal holomorphic variation operator $\Delta$
(which is roughly defined by setting $\bar\mu = 0$), 
we obtain the simple expression $\Delta J = - 2\ii \mu$. 
More generally, these formal holomorphic variations are defined (with respect to some complex coordinates) by varying the complex coordinates, but not their complex conjugates, thus locally complexifying the space.

Next, we wish to examine which of our nearby almost complex structures is integrable, assuming that the original $J$ is itself integrable. One way to characterise an integrable complex structure is that the corresponding Dolbeault operator $\bar\der : \Omega^{p,q} \ra \Omega^{p,q+1}$ squares to zero. Simple reasoning shows that the deformed Dolbeault operator is exactly $\bar\der'  = \bar\der  + \mu^a \der_a $. This squares to zero provided that $\mu$ satisfies the Maurer--Cartan equation
\begin{equation}
\label{eq:Cstr-MC}
	\bar\der \mu + \tfrac12 [\mu,\mu] = 0
\end{equation}
where the operator 
\begin{equation}
\bar\der : \Omega^{0,p}(T^{(1,0)}) \ra \Omega^{0,p+1}(T^{(1,0)})
\end{equation}
and the bracket 
\begin{equation}
[ , ] : \Omega^{0,p}(T^{(1,0)}) \times \Omega^{0,q}(T^{(1,0)}) \ra \Omega^{0,p+q}(T^{(1,0)})
\end{equation}
define the Kodaira--Spencer differential graded Lie algebra (DGLA) on $\Omega^{0,\bullet}(T^{(1,0)})$. This equation is of finite order in $\mu$ because we chose a particularly nice parameter $\mu$ to label our deformations of the starting almost complex structure. Essentially, our purpose here is to find a similar Maurer--Cartan equation for the full Hull--Strominger system, rather than merely the complex manifold condition.

To proceed, let us consider how one might construct the parameter $\mu$ which resulted in the neat equation~\eqref{eq:Cstr-MC}. Consider that, viewed as a $\GL(3,\bbC)$ structure on the tangent bundle, an almost complex structure is simply an element of the coset
\begin{equation}
\label{eq:Cstr-coset}
	\frac{\GL(6,\bbR)}{\GL(3,\bbC)}
\end{equation}
at each point of the manifold (varying smoothly). We can consider the Lie algebra decomposition
\begin{equation}
	\mathfrak{gl}(6,\bbR) \ra 
		\mathfrak{gl}(3,\bbC) 
		\oplus \Big[ (\rep{3}\otimes\rep{\bar3'}) \oplus (\rep{\bar3}\otimes\rep{3'}) \Big]_\bbR
\end{equation}
to see that a neighbourhood of the origin in $\rep{3}\otimes\rep{\bar3'}$ can be used as a local coordinate chart around the origin of~\eqref{eq:Cstr-coset}. At each point of the manifold, our tensor $\mu$ above provides precisely a point near the origin in this representation. 

We can also follow this group theoretical prescription in the case of the full $\cN=1$ geometry considered above. For simplicity, we will truncate the gauge fields and ignore the $R\wedge R$ terms (so that $H = \dd B$) for now, 
discussing their reintroduction later 
in Section~\ref{sec:gauge-fields}. 
This means that the fields considered are described by $\SO(6,6)\times\bbR^+$ generalised geometry. 
In contrast to the $\SU(3)\times\SU(3)$ structures for $\cN=2$ backgrounds discussed in Section~\ref{sec:Gen-geom}, 
the off-shell $\cN=1$ geometries require the existence of a single globally defined spinor field on $X$ which defines an $\SU(3)\times\SO(6)$ structure on the generalised tangent bundle. This is equivalent to a smoothly varying element of the homogeneous space
\begin{equation}
\label{eq:N=1coset}
	\frac{\SO(6,6)\times\bbR^+}{\SU(3)\times\SO(6)}
\end{equation}
One can then perform a decomposition of the adjoint representation of $\SO(6,6)\times\bbR^+$ to identify complex coordinates on this space, which when allowed to vary smoothly over the manifold $X$ provide us with complex coordinates on the space of structures. In~\cite{Ashmore:2018ybe} it is argued that the relevant coordinates on the space of structures can be identified with $\mu \in \Omega^{0,1}(T^{(1,0)})$ (which is $\tfrac{\ii}{2} \Delta J$ exactly as for the almost complex structure), and $x\in\Omega^{1,1}$ and $b\in\Omega^{0,2}$ which are the $(1,1)$ and $(0,2)$ parts of the holomorphic variation $\Delta(B+\ii \omega)$.

Armed with these complex coordinates on the space of $\cN=1$ structures, we can expand the superpotential in them around a supersymmetric point where $W=\delta W = 0$. As the superpotential in $\cN=1$ supergravity must be a holomorphic function of the complex scalar fields, we have that the variation of the superpotential is exactly equal to its holomorphic variation. Thus we expand
\begin{equation}
\begin{split}
W&+\Delta W = \Delta W =\int_{X} 
(H+\ii\,\dd\omega+\dd (\Delta B + \ii \, \Delta \omega)\bigl)\wedge(\Omega + \Delta\Omega) \\
&=2\int_{X}\Big(\mu^{d}\wedge\bar{\partial}x_{d}
+ \tfrac12 \mu^{d}\wedge\mu^{e}\wedge H_{de} \\
& \qquad \qquad\qquad+\mu^{d}\wedge\mu^{e}\wedge\partial_{d}x_{e}
-\tfrac{1}{2}\mu^{d}\wedge\partial_{d}\tilde{b}\Big)\wedge\Omega
\end{split}
\end{equation}
where we have suppressed the anti-holomorphic form indices. 

It is then natural to combine $\mu \in \Omega^{0,1}(T^{(1,0)})$ and $x \in \Omega^{0,1}(T^{*(1,0)})$ into $y = \mu + x \in \Omega^{0,1}(\Q')$ where $\Q'$ is the holomorphic Courant algebroid~\cite{Gualtieri:1007.3485}
\begin{equation}
\Q'\simeq T^{(1,0)}X  \oplus T^{*(1,0)}X.
\end{equation}
This algebroid is equipped with a natural holomorphic structure $\bar{D} : \Omega^{0,p}(\Q') \ra \Omega^{0,p+1}(\Q')$
\begin{equation}
\label{eq:Dbar_definition}
\begin{split}
(\bar{D}y)^{a} &=\bar{\partial}\mu^{a}, \\
(\bar{D}y)_{a} &=\bar{\partial}x_{a}
	+\ii(\partial\omega)_{ea\bar{c}}
	e^{\bar{c}}\wedge\mu^{e}.
\end{split}
\end{equation}
which is a derivation of both the holomorphic analogue of the Courant bracket and the natural pairing on $\Q'$
\begin{align}
[\cdot,\cdot]&\colon \Omega^{0,p}(\mathcal{Q}')\times\Omega^{0,q}(\mathcal{Q}') \rightarrow \Omega^{0,p+q}(\mathcal{Q}'), \\
\langle\cdot,\cdot\rangle&\colon \Omega^{0,p}(\mathcal{Q}')\times\Omega^{0,q}(\mathcal{Q}') \rightarrow \Omega^{0,p+q}(X).
\end{align}
given by
\begin{equation}
\langle y,y'\rangle=\mu^{d}\wedge x'_{d}+x_{d}\wedge\mu'^{d}.\label{eq:pairing_definition}
\end{equation}
and 
\begin{equation}
\begin{split}[y,y']_{a} & =\mu^{d}\wedge\partial_{d}x'_{a}-\partial_{d}x_{a}\wedge\mu'^{d}-\tfrac{1}{2}\mu^{d}\wedge\partial_{a}x'_{d} \\
& \qquad +\tfrac{1}{2}\partial_{a}x_{d}\wedge\mu'^{d}+\tfrac{1}{2}\partial_{a}\mu^{d}\wedge x'_{d}-\tfrac{1}{2}x_{d}\wedge\partial_{a}\mu'^{d},\\{}
[y,y']^{a} & =\mu^{b}\wedge\partial_{b}\mu'^{a}-\partial_{b}\mu^{a}\wedge\mu'^{b}.
\end{split}
\label{eq:bracket_definintion}
\end{equation}
Unlike the Lie bracket appearing in the Kodaira--Spencer DGLA, this bracket does not satisfy graded Jacobi, but rather an identity of the type
\begin{equation}
\label{eq:Courant-Jacobi}
[y,[y,y]]=-\frac{1}{3!}\partial\langle y,[y,y]\rangle,
\end{equation}
where $\der : \Omega^{0,p} \ra \Omega^{0,p}(\mathcal{Q}')$ is built from the usual Dolbeault $\der$ operator composed with the dual of the anchor map $\mathcal{Q}'\ra T^{(1,0)}$. 
As we describe below, the relation~\eqref{eq:Courant-Jacobi} becomes part of the definition of an $L_\infty$ algebra rather than a DGLA.

Returning to our expression for the superpotential expanded in the parameters $\mu$, $x$ and $b$, we see that in terms of our new parameter $y=\mu + x$ and $b \in \Omega^{0,2}$ we have the somewhat neater expression
\begin{equation}
\label{eq:W-holo-CS}
\Delta W=\int_{X}\langle y,\bar{D}y-\tfrac{1}{3}[y,y]-\partial b\rangle\wedge\Omega .
\end{equation}
The F-term conditions $W = \delta W = 0$ then (after some manipulation) take the concise form
\begin{subequations}
\label{eq:full_MC}
\begin{align}
\bar{D}y-\tfrac{1}{2}[y,y]-\tfrac{1}{2}\partial b  & =0,\label{eq:full_MC1}\\
\bar{\partial} b -\tfrac{1}{2}\langle y,\partial b\rangle +\tfrac{1}{3!}\langle y,[y,y]\rangle & =0,\label{eq:full_MC2}\\
\partial\imath_{\mu}\Omega & =0\label{eq:full_MC3}.
\end{align}
\end{subequations}
The last of these equations can be interpreted as a volume preservation condition (and has been employed as a gauge fixing condition in the Kodaira--Spencer gravity of~\cite{Bershadsky:1993cx}). 
We turn our attention to the interpretation of the first two equations below.


\subsection{$L_{\infty}$-structure}\label{sec:Linf}

Recall that an $L_\infty$-algebra is a graded vector space
\begin{equation}
\mathcal{Y}=\bigoplus_{n}\mathcal{Y}_{n},\qquad n\in\mathbb{Z},
\end{equation}
equipped with multilinear products $\ell_k$ of degree $2-k$. These are graded commutative so that
\begin{equation}
\ell_{k}(Y^{\sigma(1)},\ldots,Y^{\sigma(k)})=(-1)^{\sigma}\epsilon(\sigma;Y)\ell_{k}(Y^{1},\ldots,Y^{k}),
\end{equation}
where $(-1)^{\sigma}$ is the signature of the permutation $\sigma$ and $\epsilon(\sigma;Y)$ is the Koszul sign determined by
\begin{equation}
Y^{1}\wedge\ldots\wedge Y^{k}=\epsilon(\sigma;Y)Y^{\sigma(1)}\wedge\ldots\wedge Y^{\sigma(k)},
\end{equation}
where $Y\wedge Y^{'}=(-1)^{YY'}Y'\wedge Y$. The $\ell_k$ must also satisfy the generalised Jacobi identities. 
We refer the reader to e.g.~\cite{Hohm:2017pnh,Jurco:2018sby} for full details of these constructions and some of their applications to field theories in physics.

Here we note that in the field theory construction, the degree two elements have the form of the field equations, the degree one elements are the fields and the degree zero elements are the gauge transformations, while the tower of elements of negative degree form the tower of successive gauge transformations of the gauge transformations. However, the construction can also be phrased in terms of deformation theory. There, the field equation becomes the Maurer--Cartan equation, the fields become the deformations, the gauge transformations become the symmetries, while the lower degree elements are ``higher" symmetries. The field equation, or Maurer--Cartan equation, takes the form
\begin{equation}
\begin{aligned}
\mathcal{F}(Y) &=\sum_{n=1}^{\infty}\frac{(-1)^{n(n-1)/2}}{n!}\ell_{n}(Y^{n}) \\
&=\ell_{1}(Y)-\tfrac{1}{2}\ell_{2}(Y,Y)-\tfrac{1}{3!}\ell_{3}(Y,Y,Y)+\cdots = 0
\end{aligned}
\end{equation}
where $Y\in\mathcal{Y}_{1}$, while the gauge transformation or symmetry of the system, takes the form
\begin{equation}
\begin{aligned}
\delta_{\Lambda}Y &=\ell_{1}(\Lambda)+\ell_{2}(\Lambda,Y)-\tfrac{1}{2}\ell_{3}(\Lambda,Y,Y)\,-\\
&\kern.5cm-\,\tfrac{1}{3!}\ell_{4}(\Lambda,Y,Y,Y)+\cdots,
\end{aligned}
\end{equation}
where $\Lambda\in\mathcal{Y}_{0}$.

For the heterotic system~\eqref{eq:full_MC}, the relevant $L_\infty$-algebra is built from the graded vector space with
\begin{equation}
\label{eq:het-Yn}
	\mathcal{Y}_n = \Omega^{0,n}(\Q') \oplus \Omega^{0,n+1}
\end{equation}
while the multilinear products $\ell_k$ on elements $Y = (y, b)$ of definite degree are given by
\begin{equation}\label{eq:Linfty_products}
\begin{split}\ell_{1}(Y) & :=(\bar{D}y+\tfrac{1}{2}(-1)^Y\partial b,\bar{\partial}b),\\
\ell_{2}(Y,Y') & :=([y,y'],\tfrac{1}{2}(\langle y,\partial b'\rangle+(-1)^{1+YY'}\langle y',\partial b\rangle)),\\
\ell_{3}(Y,Y',Y'') & :=\tfrac{1}{3}(-1)^{Y+Y'+Y''}(0,\langle y,[y',y'']\rangle\,+\\
& \kern1cm+\,(-1)^{Y(Y'+Y'')}\langle y',[y'',y]\rangle\,+\\
& \kern1cm+\,(-1)^{Y''(Y+Y')\,}\langle y'',[y,y']\rangle),\\
\ell_{k\geq4} & := 0.
\end{split}
\end{equation}
As our highest degree non-trivial bracket takes three arguments this is an $L_3$-algebra. The Maurer--Cartan equation arising from this is then
\begin{equation}
\begin{aligned}
\mathcal{F}(Y) &=(\bar{D}y-\tfrac{1}{2}\partial b -\tfrac{1}{2}[y,y],\bar{\partial} b -\tfrac{1}{2}\langle y, b \rangle\,+\\
&\kern.5cm+\,\tfrac{1}{3!}\langle y,[y,y]\rangle) = 0.
\end{aligned}
\end{equation}
Remarkably, this encapsulates precisely equations~\eqref{eq:full_MC1} and~\eqref{eq:full_MC2}. 
In other words, the F-term conditions for a supersymmetric Minkowski vacuum are equivalent to
\begin{equation}
\mathcal{F}(Y) = 0, \qquad \partial\imath_\mu \Omega=0.
\end{equation}
The second condition here can be absorbed into the Maurer--Cartan equation by restricting the graded vector spaces to the sheaves of sections where this condition holds.

We also note that there is a natural construction~\cite{Roytenberg:1998vn} of an $L_3$-algebra associated to any Courant algebroid, with the degree zero elements as its sections and the degree $-1$ elements the smooth functions on the base manifold. This construction can be adapted to the holomorphic case, giving an $L_3$-algebra on the two-term complex
\begin{equation}
\label{eq:holo-L3}
	\mathcal{O}_X \stackrel{\der}{\longrightarrow} \mathcal{E} .
\end{equation}
where $\mathcal{E}$ is the sheaf of holomorphic sections of $\Q'$. One can then consider the Dolbeault resolution of this complex (see Figure~\ref{fig:resolution})  
\begin{figure*}[h]
\begin{center}
\begin{equation*}
\kern2.2cm\begin{tikzpicture}[scale=1.75,baseline=(current bounding box.center)] 
\node (Az) at (-2.5,1) {$0$}; 
\node (Ah) at (-1.3,1) {$\mathcal{O}_X$}; 
\node (A0) at (0,1) {$\mathcal{C}^\infty (\bbC)$}; 
\node (A1) at (1.5,1) {$\Omega^{0,1}$}; 
\node (A2) at (3.2,1) {$\Omega^{0,2}$}; 
\node (A3) at (4.5,1) {$$}; 
\node (Bz) at (-2.5,0) {$0$}; 
\node (Bh) at (-1.3,0) {$\mathcal{E}$}; 
\node (B0) at (0,0) {$\Q'$}; 
\node (B1) at (1.5,0) {$\Omega^{0,1}(\Q')$}; 
\node (B2) at (3.2,0) {$\Omega^{0,2}(\Q')$}; 
\node (B3) at (4.5,0) {$$}; 
\path[->,font=\scriptsize] 
(Az) edge node[above]{$$} (Ah)
(Ah) edge node[above]{$\iota$} (A0)
(A0) edge node[above]{$\bar\der$} (A1)
(A1) edge node[above]{$\bar\der$} (A2)
(A2) edge node[above]{$\bar\der$} (A3)
(Bz) edge node[above]{$$} (Bh)
(Bh) edge node[above]{$\iota$} (B0)
(B0) edge node[above]{$\bar{D}$} (B1)
(B1) edge node[above]{$\bar{D}$} (B2)
(B2) edge node[above]{$\bar{D}$} (B3)
(Ah) edge node[left]{$\der$} (Bh)
(A0) edge node[left]{$\der$} (B0)
(A1) edge node[left]{$\der$} (B1)
(A2) edge node[left]{$\der$} (B2);
\end{tikzpicture}
\end{equation*}
  \caption{\label{fig:resolution} The Dolbeault resolution of~\eqref{eq:holo-L3}.
  }
 \end{center}
\end{figure*}
and see that the resulting total complex is that of our $L_3$-algebra~\eqref{eq:het-Yn}. One can go on to see that the relation between~\eqref{eq:het-Yn} and~\eqref{eq:holo-L3} is a quasi-isomorphism. This provides a natural mathematical construction of~\eqref{eq:het-Yn} from the holomorphic Courant algebroid.

\subsection{Restoring the gauge fields}
\label{sec:gauge-fields}

In the above, we truncated the gauge fields of heterotic supergravity for simplicity. We now comment on how we can include them in the construction. The first step is to again simplify the problem by treating the tangent bundle connection $\Theta$ appearing in~\eqref{eq:het-H} as an additional degree of freedom (rather than fixing it to be a specific connection determined by the physical fields as required by supersymmetry). In this way, we can treat it as an additional gauge field, as in the infinitesimal treatments of~\cite{delaOssa:2014cia,Anderson:2014xha}, so that the Bianchi identity can be written as $\dd H = \dd B + \tfrac{\alpha'}{4} F \wedge F$. 

The necessary modifications to our construction then follow simply by replacing $\Q' \simeq T^{(1,0)} \oplus T^{*(1,0)}$ by an enhanced holomorphic Courant algebroid (also recently discussed in~\cite{Garcia-Fernandez:2018emx,Garcia-Fernandez:2018ypt} with the nomenclature `holomorphic string algebroid')
\begin{equation}\label{eq:Q}
\Q \simeq T^{(1,0)}X  \oplus \End V \oplus T^{*(1,0)}(X).
\end{equation}
We refer the reader to~\cite{Ashmore:2018ybe} for the details of the modifications to the pairing, bracket and holomorphic structure needed for this bundle. The main statement is that having made these modifications, the exact same form of equations~\eqref{eq:W-holo-CS} and~\eqref{eq:full_MC} and those of Section~\ref{sec:Linf} hold for the case including the gauge bundle.

Strictly though, we have solved a different problem to the original moduli problem, as we introduced extra degrees of freedom to the theory. The true moduli space will then be a subspace of the moduli space found in this way, on which the additional `gauge field' $\Theta$ is fixed to be the Hull connection (see the discussion in~\cite{delaOssa:2014cia}). How best to describe these additional constraints remains an open problem.

\bibliography{allbibtex}

\bibliographystyle{prop2015}

\end{document}